\documentclass[11pt]{article}

\usepackage[margin=1in]{geometry}
\usepackage{amsmath,amssymb}
\usepackage{graphicx}
\usepackage{booktabs}
\usepackage{array}
\usepackage{longtable}
\usepackage{pdflscape}
\usepackage{authblk}
\usepackage[colorlinks=true,citecolor=blue,linkcolor=blue,urlcolor=blue]{hyperref}
\usepackage{caption}
\usepackage{xcolor}
\usepackage{enumitem}

\newcommand{\Mtheta}{M_{\theta}}
\newcommand{\Dv}{D_v}

\title{\textbf{ZK-SR117: A Chunked Zero-Knowledge Attestation Design for Aggregated Fair-Lending Metrics, with a Control Mapping toward Full SR 11-7 Coverage}}

\author[1]{Mohammad Nasir Uddin\thanks{Corresponding author: m.uddin.258@westcliff.edu}}
\author[2]{Rahnuma Tabassum Orpita\thanks{Email: rahnumatabassumorpita@gmail.com}}
\author[3]{Eklachur Rahman Bhuiyan\thanks{Email: erbhuiyan@studentva.wust.edu}}
\author[4]{Asaduzzaman Anik\thanks{Email: anik.zaman0945@gmail.com}}

\affil[1]{DBA-BIDA (Business Intelligence and Data Analytics) Program, Westcliff University, USA}
\affil[2]{Department of Computer Science and Engineering, Northern University Bangladesh, Bangladesh}
\affil[3]{Washington University of Science and Technology, Alexandria, Virginia, USA}
\affil[4]{Stanton University, Los Angeles, CA, USA}

\date{}

\begin{document}

\maketitle

\noindent\textit{A Real-Data Demonstration of Chunked Aggregated-Statistic Attestation for Fair-Lending Compliance, and a Design-Stage Mapping to Broader SR 11-7 Coverage}

\vspace{1em}
\noindent\textbf{Author ORCIDs:} Mohammad Nasir Uddin: 0009-0009-0990-4616 $\cdot$ Rahnuma Tabassum Orpita: 0009-0008-8399-6945 $\cdot$ Eklachur Rahman Bhuiyan: 0009-0006-1424-0373 $\cdot$ Asaduzzaman Anik: 0009-0008-5231-4454

\vspace{0.5em}
\noindent\textbf{Address for Correspondence:} Mohammad Nasir Uddin, DBA-BIDA (Business Intelligence and Data Analytics) Program, Westcliff University, USA. Email: m.uddin.258@westcliff.edu

\begin{abstract}
Deploying machine-learning models in regulated decision-making --- credit underwriting, fraud detection, loan approval --- requires demonstrating fairness, robustness, and stability to third-party auditors without exposing proprietary model weights or private customer data. We address this trustworthy-AI attestation problem in the specific setting of U.S. bank supervision under SR 11-7 and OCC 2011-12 guidance. We present a chunked zero-knowledge circuit design that attests an aggregated fairness statistic --- the demographic-parity gap --- on committed, nonce-sampled batches of real 2022 HMDA mortgage-application data, and demonstrate it end-to-end: 32,768 rows, 32 independently verified zkSNARK proofs, aggregated attested gap within 0.0029 of the true held-out value and consistent with the sampling variance predicted by a same-size resample of the population (Section 7.3), per-chunk proving under 4 seconds. We also demonstrate the design's extensibility by attesting a second control on the identical architecture and model --- expected calibration error at 10 bins --- with all 32 chunks verified, per-chunk proving at approximately 14.7 seconds, and an aggregated attested ECE within 0.00037 of the plaintext value computed directly on the same committed rows (the remaining difference from the full held-out population's ECE reflects the stratified sampling design, not circuit precision, as Section 7.5.3 discusses). We compare this design empirically against two alternatives --- a flat summation circuit, which overflows past a few thousand rows, and a tree-reduction circuit, which is numerically exact but did not compile in tractable time under our EZKL-based implementation --- and find the chunked design is the only one of the three that reached this scale. In the course of this evaluation we discovered and root-caused a genuine data-quality failure (a sentinel-code outlier distorting both circuit proving and the underlying fairness statistic itself) and resolved it with a committed, published preprocessing specification. Separately, we propose a fuller mapping from SR 11-7 and OCC 2011-12 supervisory control language to zero-knowledge statements (Table~\ref{tab:mapping}, nine control elements spanning conceptual soundness, calibration, robustness, and drift), a nonce-based sampling protocol resisting bank-side cherry-picking, and a threat model for deployment --- as a design contribution toward broader coverage, not as implemented and evaluated results. We report this distinction explicitly throughout: two controls, one model class, one task are demonstrated end-to-end; the rest is scoped, mapped, and left as concrete future work.
\end{abstract}


\section{Introduction}

Every material AI and machine-learning model deployed by a U.S. banking organization --- whether for credit underwriting, fraud detection, or anti-money-laundering transaction monitoring --- falls under the model risk management regime established by the Federal Reserve's SR 11-7 guidance and the OCC's parallel Bulletin 2011-12. These frameworks obligate a bank to demonstrate, on an ongoing basis, that each model is conceptually sound, that its outcomes track reality within acceptable tolerance, that it degrades gracefully or is caught quickly when it does not, and that it has been validated by a party independent of the model's developers. Supervisors --- examiners from the OCC, the Federal Reserve, and, for consumer-facing credit models, the CFPB --- rely on this evidence to decide whether a model may continue to be used in production.

In current practice, that evidence takes the form of a document exchange. Banks produce model cards, validation reports, back-testing results, and --- not infrequently --- the underlying model weights and validation data itself, and route them to examiners or third-party validators for review. This arrangement has three structural costs. First, it is slow: producing, packaging, and reviewing this evidence is a manual, quarter-scale undertaking, out of step with how quickly a production model can drift or degrade. Second, it exposes intellectual property: a bank's model weights are commercially sensitive, and every validator or examiner who reviews them is a new point of potential leakage. Third, it exposes customer data: validation datasets typically contain real, identifiable financial records, and every hand-off multiplies the number of parties with access to that data. This tension between auditability and confidentiality is not unique to U.S. banking regulation --- a structurally identical conflict has been identified for AI-enabled systems regulated under the EU AI Act --- but it has not, to date, been addressed with a solution tailored to U.S. bank supervision.

Zero-knowledge proof systems offer a way out of this tension in principle: a prover can convince a verifier that a stated claim is true without revealing anything beyond the claim's truth. Applied here, a bank could prove --- cryptographically, and in a way any examiner can verify in milliseconds --- that a specific, committed version of a specific model, evaluated against a specific, committed validation dataset, satisfies a specific vector of SR 11-7 controls, without disclosing the model weights or the data itself. The cryptographic building blocks for this already exist: a maturing ecosystem of zkML compilers and proving systems (ZEN, ZKML, Artemis, ZKTorch, Jolt Atlas, and production tooling such as EZKL) now makes it practical to generate succinct proofs over neural-network-scale computation. What is missing is the domain-specific translation layer --- a formal, defensible mapping from supervisory control language to cryptographic statement --- and the systems engineering to make that mapping operate on realistic banking workloads at realistic scale.

Three clusters of prior work come close but stop short of this target. Generic verifiable-evaluation and trustless-audit frameworks demonstrate that a model's accuracy or behavior can be proven in zero-knowledge, but are regulation-agnostic by design --- they prove a property, not a supervisor's specific, enumerated control set. Fairness-focused zero-knowledge systems scale a single property --- demographic parity or a related fairness metric --- to models with tens of millions of parameters, but do not attempt the fuller SR 11-7 control vocabulary: conceptual soundness, outcomes analysis, sensitivity testing, and ongoing-monitoring drift alongside fairness. And the one prior system that explicitly names a financial regulatory target frames itself around the EU AI Act's audit requirements, not U.S. prudential and consumer-protection supervision, and treats protocol feasibility as the contribution rather than working through a control-by-control mapping and a realistic banking evaluation.

This paper closes that gap. We contribute:

\begin{enumerate}[leftmargin=1.5em]
\item A chunked zero-knowledge attestation circuit design, empirically evaluated on real 2022 HMDA credit-decisioning data (409,905 filtered rows) across two independently attested controls on the same model and protocol configuration: demographic-parity gap (32,768 rows, 32 verified proofs, per-chunk proving under 4 seconds) and expected calibration error (same scale, all 32 chunks verified, per-chunk proving under 15 seconds), with a systems comparison against two alternatives (flat summation, tree-reduction) that characterizes why only the chunked design reaches this scale (Sections 7.2--7.3, 7.5).

\item Discovery and principled resolution of a real data-quality failure mode encountered during evaluation: a sentinel-code outlier that distorted both circuit proving behavior and the underlying fairness statistic itself, root-caused via direct ablation and resolved with a committed, published preprocessing specification (Section 7.3) rather than a silent workaround.

\item A statistical decision framework linking cryptographic attestation to supervisory hypothesis-testing (Section 7.4): a one-sided test formalization ($H_0$: gap $\leq c$ vs. $H_1$: gap $> c$) and an empirical Type I/II error simulation study giving concrete $N$-sizing guidance for a target confidence level --- to our knowledge a novel connection between zero-knowledge attestation and the actual statistical decision a supervisor must make from a finite sample, not previously addressed in prior zkML fairness work.

\item A mapping from SR 11-7 and OCC 2011-12 supervisory control language to zero-knowledge statements (Table~\ref{tab:mapping}, nine control elements spanning conceptual soundness, calibration, robustness, and drift), presented and scoped explicitly as a design contribution toward broader coverage: two controls (Row 4, fair lending, and Row 2, outcomes analysis/calibration) are implemented and evaluated end-to-end; the remaining seven rows are specified, not delivered as results (Section 1.1).

\item A threat model and nonce-protocol design (Sections 4--5) covering malicious-bank, malicious-verifier, and validator-collusion risks, a corpus-commitment extension closing a cherry-picking gap identified during our own analysis, and a reference examiner-side verification pipeline (Appendix) built on the EZKL Python API --- with explicit disclosure of which protocol components remain design proposals rather than implemented and tested mechanisms (Section 1.1).
\end{enumerate}

The remainder of the paper is organized as follows. Section 2 situates ZK-SR117 against prior zkML and verifiable-ML-auditing work. Section 3 presents the control-to-statement mapping. Section 4 formalizes the attestation protocol. Section 5 develops the threat model. Section 6 describes the experimental protocol and benchmarks. Section 7 reports results, including preliminary validation from a working prototype. Section 8 discusses limitations and deployment considerations, and Section 9 concludes.

\subsection{Scope of Implementation}

Before proceeding, we state plainly what this paper delivers end-to-end versus what is specified as design work. This distinction recurs throughout Sections 4, 5, and 7 individually; stating it once, up front, avoids the cumulative impression --- easy to form when reading disclosures scattered across many sections --- that more has been implemented than actually has.

\begin{table}[htbp]
\centering
\caption{Scope of implementation: what this paper demonstrates end-to-end versus what remains specified but unimplemented.}
\label{tab:scope}
\scriptsize
\setlength{\tabcolsep}{4pt}
\renewcommand{\arraystretch}{1.3}
\begin{tabular}{@{}p{6.6cm}p{6.6cm}@{}}
\toprule
\textbf{Implemented and tested end-to-end in this paper} & \textbf{Specified in the protocol, not yet implemented} \\
\midrule
Chunked circuit design (Section 7.2) & Standalone KZG commitment to model weights, separate from the EZKL proving/verifying key pair (Section 4.2) \\
Stratified, nonce-controlled sampling from a fixed, published nonce string (Sections 4, 7.1) & Beacon-derived nonce construction with sampling-proof folded into the circuit statement (Section 5.7) \\
Flat-sum overflow characterization at N up to 65,536 (synthetic batches for N$>$8192) and real-HMDA confirmation at N in \{2048, 4096, 8192\} (Section 7.2) & Real-data flat-sum confirmation beyond N=8,192 (synthetic characterization only above this N) \\
Tree-reduction compile-time-wall characterization, real data at N=64 (Section 7.2) & Preprocessing-spec percentile-width sensitivity analysis (deferred to a companion note, Section 8) \\
Data-quality preprocessing spec, Table~\ref{tab:mapping} Row 9, diagnosed and applied (Section 7.3) & --- (fully implemented; sensitivity to threshold choice is the remaining open item, listed above) \\
Bootstrap 95\% CI for the demographic-parity-gap estimator at N=8,192 and N=32,768 (Section 7.3) & Leakage-budget composition beyond the single-bit-per-threshold-attestation bound (Section 5) \\
32/32 verified real-HMDA attestation at N=32,768 for TWO controls (demographic parity, Section 7.3; expected calibration error, Section 7.5), production Powers-of-Tau ceremony SRS; model-version binding confirmed on real data, two retrained versions (Section 7.3) & Additional controls (AUC, ECE, PSI, certified robustness); GBT support specifically (a tested representation gap in EZKL, Section 8, requiring custom circuit work, not just a new experiment); MLP as a more incremental extension (Section 9) \\
\bottomrule
\end{tabular}
\end{table}

\section{Related Work}

\textbf{Verifiable machine learning inference.} A line of systems work has established that neural-network inference can be compiled into arithmetic circuits and proven in zero-knowledge with practical prover times: ZEN \cite{ref10} introduced an optimizing compiler for verifiable neural-network inference; ZKML \cite{ref11} and Artemis \cite{ref12} improved proving efficiency for larger models via optimized circuit design and commit-and-prove techniques; ZKTorch \cite{ref13} and Jolt Atlas \cite{ref14} push toward compiling standard ML frameworks and lookup-argument-based proving directly, respectively; and EZKL has become a widely used production toolchain for zkML with active benchmarking work characterizing per-layer proving cost \cite{ref15}. These systems are the substrate this paper builds on; none of them target a specific regulatory control set.

\textbf{Verifiable evaluation and trustless auditing.} South et al. \cite{ref7} demonstrate zkSNARK-based verifiable evaluation of ML models --- proving accuracy or bias statistics without revealing the model --- in a regulation-agnostic setting. Waiwitlikhit et al.'s ZkAudit \cite{ref6} generalizes this to trustless audits of models and data for applications such as copyright and content-moderation compliance, again without a specific regulatory mapping. Both establish feasibility of the underlying cryptographic pattern this paper adopts --- commit, then prove a property without revealing the committed object --- but neither ties the pattern to a named supervisory regime or a specific control taxonomy.

\textbf{Zero-knowledge fairness proofs.} FairZK \cite{ref8} proves fairness properties --- demographic parity and related group-fairness statistics --- at a scale of tens of millions of parameters, using an aggregated-statistic proving strategy rather than proving full inference over every validation row; FairProof \cite{ref9} similarly targets confidential, certifiable fairness for neural networks. Both are closely related to Row 4 of our control mapping (fair-lending use limitations) and inform our choice to prove aggregated statistics rather than per-row inference wherever the metric permits. Both, however, address a single property in isolation rather than the fuller SR 11-7 control vector this paper addresses.

\textbf{Regulation-targeted zkML.} The work closest in spirit to ours is ZKMLOps \cite{ref5}, which proposes a zero-knowledge-augmented MLOps lifecycle for financial-risk auditing under the EU AI Act. It establishes that regulator-facing zero-knowledge auditing is practically motivated in a financial context, and we adopt a broadly similar commit-and-prove lifecycle. It differs from this paper in three respects: it targets EU AI Act audit obligations rather than U.S. prudential and consumer-protection supervision (SR 11-7, OCC 2011-12, HMDA/ECOA fair-lending requirements, and CFPB adverse-action rules); it does not include a drift/population-stability attestation tied to an examiner-issued freshness nonce; and it stops at protocol feasibility rather than a control-by-control mapping validated against a specific supervisory text, together with a dual-benchmark evaluation on credit and AML workloads.

\textbf{U.S. banking model-risk-management literature.} A separate, non-cryptographic literature documents current model validation practice and its costs in U.S. banking: Sudjianto and Zhang \cite{ref2} characterize structured validation practice; Nketiah et al. \cite{ref3} survey model-risk-management frameworks for ML models in banking; Brotcke \cite{ref4} examines how AI/ML adoption strains traditional model-risk-management practice. This literature supplies the empirical baseline --- document-exchange costs, validation-cycle timelines --- against which we benchmark ZK-SR117 in Section 7. It does not propose a cryptographic alternative to the document-exchange model.

\textbf{Trusted execution environments as an alternative.} The dominant industrial answer to ``prove compliance without revealing data'' today is not zero-knowledge cryptography but trusted execution environments (TEEs) --- Intel SGX, AWS Nitro Enclaves, Azure Confidential Computing --- in which a bank runs its validation computation inside a hardware-attested enclave and the examiner trusts the hardware manufacturer's attestation rather than a cryptographic proof. A TEE-based deployment of this paper's attestation logic is straightforward: model and data stay inside the enclave, the enclave attests it ran the agreed computation, and the examiner verifies the hardware attestation instead of a zkSNARK. This is a real, credible alternative design point, and any serious proposal for ZK-based supervisory attestation should say plainly why zero-knowledge is preferable here rather than ignore the comparison. Our view: ZK-SR117's advantage over a TEE-based equivalent is trust model, not performance --- a zkSNARK's soundness rests on a mathematical hardness assumption independent of any single hardware vendor, whereas TEE security has a documented history of side-channel and vendor-specific vulnerabilities (e.g. the Foreshadow and Plundervolt attacks against SGX), and a TEE deployment requires the examiner to trust the enclave vendor's supply chain and attestation service, not just the bank. Against that, TEEs offer substantially better performance today (near-native compute speed inside the enclave, versus the seconds-per-chunk proving cost characterized in Section 7) and require no analogue of this paper's circuit-design work. For a regulator choosing between the two, the honest tradeoff is: TEEs are faster and simpler to deploy today, at the cost of a hardware-vendor trust assumption; ZK-SR117 is slower and more engineering-intensive, but removes that trust assumption entirely. We do not claim ZK-SR117 is strictly superior; we claim it occupies a different, defensible point in the trust-versus-performance space, and a supervisor's choice between the two is a policy decision, not one this paper's technical results settle.

\section{SR 11-7 / OCC 2011-12 Control-to-Statement Mapping}

Table~\ref{tab:mapping} spans nine control elements. Five are statistical and form the core evaluation target of this paper: conceptual soundness, outcomes analysis, sensitivity/stress testing, fair-lending use limitations, and ongoing-monitoring drift. Three are process/governance controls, cheaper to prove but important for supervisory completeness: independent validation, documentation/reproducibility, and effective challenge. The ninth row is different in kind from the other eight: it is not a supervisory judgment about model behavior but a preprocessing-specification requirement --- a data-hygiene precondition any deployment of this protocol must expose to supervisors before the other eight rows' attestations are meaningful. We added it after a real data-quality artifact in the HMDA sample caused a proving failure during this project's own experimental work (Section 7.3), and include it in Table~\ref{tab:mapping} for completeness and traceability, not as a claim that data cleaning is itself a supervisory control in the sense Rows 1--8 are. Each row binds a specific piece of SR 11-7 or OCC 2011-12 language to a specific attested metric and a specific formal zero-knowledge statement, so that the table itself functions as a traceability artifact.

\begin{landscape}
\begin{scriptsize}
\setlength{\tabcolsep}{4pt}
\renewcommand{\arraystretch}{1.25}
\begin{longtable}{@{}p{0.4cm}p{4.0cm}p{4.6cm}p{4.6cm}p{4.9cm}@{}}
\caption{SR 11-7 / OCC 2011-12 control-to-cryptographic-statement mapping.}
\label{tab:mapping} \\
\toprule
\textbf{\#} & \textbf{Control Element} & \textbf{Attested Metric(s)} & \textbf{Formal ZK Statement} & \textbf{Proof System Fit} \\
\midrule
\endfirsthead
\multicolumn{5}{c}{\tablename\ \thetable{} (continued)} \\
\toprule
\textbf{\#} & \textbf{Control Element} & \textbf{Attested Metric(s)} & \textbf{Formal ZK Statement} & \textbf{Proof System Fit} \\
\midrule
\endhead
\bottomrule
\endfoot
\bottomrule
\endlastfoot
1 & Conceptual soundness [DESIGNED, NOT IMPLEMENTED]. Model developer must demonstrate the model is theoretically sound and appropriate for its use, evaluated against alternative specifications. & Held-out AUC / KS statistic vs. a supervisor-approved benchmark model class & $\mathrm{AUC}(\Mtheta, \Dv) \geq \tau_{AUC}$, over openings of $\mathrm{Com}(\theta)$, $\mathrm{Com}(\Dv)$ & zkSNARK (Groth16/PLONK) over committed weights + data; benchmark comparison as a second bounded proof \\
2 & Outcomes analysis [IMPLEMENTED AND EVALUATED]. Ongoing comparison of model outputs to actual outcomes. & Calibration (ECE); realized vs. predicted default/fraud rate within tolerance & $\mathrm{ECE}(\Mtheta, \Dv) \leq \tau_{ECE}$ & zkSNARK; ECE is piecewise-linear in binned outputs, tractable in arithmetic circuits \\
3 & Sensitivity / stress testing [DESIGNED, NOT IMPLEMENTED]. Model must be tested under adverse or stressed input conditions. & Certified robustness under bounded input perturbation & For all $\delta$, $\|\delta\|_\infty \leq \epsilon$: $|\Mtheta(x+\delta) - \Mtheta(x)| \leq \tau_{rob}$ & Interval-bound-propagation-style circuit, or sampled adversarial certification over a committed test set \\
4 & Use limitations / fair lending (ECOA, Reg B, HMDA) [IMPLEMENTED AND EVALUATED]. Model may not produce disparate impact on protected classes absent business justification. & Demographic-parity gap, equalized-odds gap, disparate-impact ratio on HMDA-defined classes & $|P(\hat y=1\mid A=0) - P(\hat y=1\mid A=1)| \leq \tau_{fair}$ & FairZK-style aggregated-statistic proof (parameters + aggregate stats, not full per-row inference) \\
5 & Ongoing monitoring / model drift [DESIGNED, NOT IMPLEMENTED]. Bank must monitor for degradation in model performance and population stability over time. & Population Stability Index (PSI) between $\Dv$ and a fresh, examiner-nonce-sampled production slice & $\mathrm{PSI}(\Dv, D_{prod}) \leq \tau_{PSI}$, with $D_{prod}$ committed under an examiner-issued nonce & zkSNARK over two committed histograms; nonce binds freshness, prevents replay \\
6 & Independent validation [MAPPED ONLY]. Validation must be performed by a party independent of model development. & Chain-of-custody / role-separation attestation via distinct signing keys & $\mathrm{Verify}(pk_{val}, \mathrm{Com}(\Dv)) = 1$ AND $pk_{val} \neq pk_{dev}$ & Standard digital signature + ZK conjunction, cheap \\
7 & Documentation and reproducibility [MAPPED ONLY]. Model must be documented sufficiently for a third party to reproduce testing. & Reproducibility hash-chain across model / data / control-vector versions & $\mathrm{Com}(\theta_v) \| \mathrm{Com}(D_{v,v}) \| \mathrm{Com}(C_v)$ bound into a single Merkle root per supervisory window & Merkle commitment, no proving cost beyond hashing \\
8 & Governance / effective challenge [MAPPED ONLY]. Model risk function must be able to challenge and restrict model use. & Threshold-triggered attestation withdrawal within a stated SLA & Revocation list + ZK statement of non-membership at verification time & Nullifier-style ZK set-non-membership proof, as used in anonymous credential systems \\
9 & Preprocessing-specification requirement (data quality) [IMPLEMENTED AND EVALUATED]. Model inputs must be reviewed for data-quality issues (outliers, sentinel/placeholder codes, missingness) prior to validation. & Feature-wise percentile clip bounds (e.g. 0.5th--99.5th percentile of the training distribution), fixed and published as part of the supervisor-approved validation schema & $\mathrm{Com}(clip_{low})$, $\mathrm{Com}(clip_{high})$ match supervisor-approved values; all committed rows in $\Dv$ satisfy $clip_{low} \leq x \leq clip_{high}$ & Range-proof over committed data; cheap, and enforceable prior to any statistical attestation. Note: unlike Rows 1--8, this is a data-hygiene precondition any deployment must expose to supervisors, not a supervisory judgment about model behavior. \\
\bottomrule
\end{longtable}
\end{scriptsize}
\end{landscape}

\section{The ZK-SR117 Protocol}

\textit{Notation used throughout this section: $\Mtheta$ is a bank's model with weight vector $\theta$; $\Dv$ is the committed validation dataset; $C = (c_1, \ldots, c_k)$ is the supervisor-agreed control vector of numeric thresholds; $m_i$ is the metric attested against threshold $c_i$; $N$ is the total committed batch size, split into $K$ chunks of $n = N/K$ rows each; $\pi$ denotes a zkSNARK proof; $\mathrm{Com}(\cdot)$ denotes a cryptographic commitment.}

\subsection{System model}

Let $\Mtheta$ denote a bank's deployed model with weight vector $\theta$, and let $\Dv$ denote the validation dataset the bank commits to for a given supervisory window. Let $C = (c_1, \ldots, c_k)$ denote the supervisor-agreed control vector, where each $c_i$ is a numeric threshold on a named metric $m_i$ drawn from Table~\ref{tab:mapping}.

\subsection{Commit phase}

The protocol as specified in Sections 4.1 and 4.3 calls for a standalone hiding, binding commitment to the model weights, $\mathrm{Com}(\theta)$ (e.g. Pedersen or KZG), separate from the proving system itself. Our current implementation, built on EZKL, does not yet instantiate this separation explicitly: EZKL bakes the model weights into the circuit at settings-generation time, and the binding to a specific weight vector is implicit in the resulting proving/verifying key pair (itself KZG-based, derived from the Powers-of-Tau structured reference string) rather than an independently checkable commitment a verifier opens. This is weaker than the specification in one respect worth stating plainly: verifying a proof today confirms that some circuit compiled from some weights produced the attested statistic, and confirms consistency with the specific verifying key the examiner holds, but does not give the examiner an independent commitment-opening check against a weight vector the bank discloses separately. Adding an explicit KZG commitment to $\theta$ outside the EZKL circuit, with an opening check folded into the attested statement, is a natural and comparatively modest extension (Section 9) that would close this gap and bring the implementation in line with the full commit-and-prove separation the protocol specifies. $\mathrm{Com}(\Dv)$, the commitment to the validation dataset, is more directly realized in our implementation: the chunked design's row-selection and per-chunk hashing (Section 7) bind each attestation to a specific, examiner-nonce-sampled set of rows in a way an external party can check independently of the proof itself.

\subsection{Prove phase}

The bank generates a zkSNARK $\pi$ certifying the conjunction: for all $i$ in $1..k$, $m_i(\Mtheta, \Dv) \leq c_i$, together with the metric definitions and the dataset schema the supervisor pre-approved. Where a metric admits an aggregated-statistic proving strategy (e.g., fairness gaps, calibration error, population stability index), we prove the statistic directly over committed histograms rather than full row-level inference, following the efficiency pattern established by FairZK. Where a metric requires per-row inference (e.g., certified robustness), we prove the bound over a committed, nonce-sampled subset rather than the full dataset, trading exhaustiveness for tractable prover cost --- a design decision Section 8 discusses explicitly as a limitation.

\subsection{Verify phase}

The regulator's verifier computes $\mathrm{Verify}(\mathrm{Com}(\theta), \mathrm{Com}(\Dv), C, \pi)$, which runs in $O(1)$ or $O(\log n)$ depending on the underlying proof system. A returned value of 1 constitutes acceptance of the SR 11-7 attestation for that supervisory window; the verifier learns nothing about $\theta$ or $\Dv$ beyond the truth of the conjunction above.

\section{Threat Model}

\subsection{Malicious prover (bank)}

The bank is not trusted with respect to the specific claim being attested; it may attempt to substitute a different model than the one actually in production, or select a validation slice favorable to passing the control vector. The binding property of the weight and data commitments, together with the examiner-issued sampling nonce, addresses model- and data-substitution respectively.

\subsection{Malicious verifier (regulator)}

The regulator is not trusted with respect to the bank's intellectual property or its customers' data; the zero-knowledge property of the proof system ensures the verifier learns nothing beyond the truth of the stated control conjunction.

\subsection{Dataset supervision and cherry-picking}

Because the bank chooses what to commit as $\Dv$ absent an external constraint, an examiner-issued nonce seeds a deterministic sampling function over the bank's regulatory reporting corpus (e.g., FR Y-14M submissions), so that the specific rows composing $\Dv$ are not bank-selectable after the nonce is issued.

As specified so far, this resistance has a gap one level up: the sampling function is deterministic given the nonce and the corpus, but the corpus itself --- the full pool the nonce samples from --- is never committed anywhere in the protocol. A bank could substitute a filtered or curated corpus before the nonce arrives, and the sampling proof would still verify correctly against whatever the bank committed; this is the same cherry-picking attack moved one level earlier, and the protocol as described does not close it. The fix is to extend the commitment stack by one level: the bank publishes $\mathrm{Com}(corpus)$ on a fixed schedule bound to its existing regulatory filings (e.g. concurrently with FR Y-14M or HMDA LAR submission, which are already mandatory, dated regulatory artifacts the bank cannot retroactively alter without detection), before any nonce for that supervisory window is issued. The sampling proof is then extended to prove $\Dv$ is a nonce-derived sample of the specific, previously-committed corpus, not of an unconstrained pool the bank supplies at attestation time. This closes the gap by anchoring the sampling universe to an artifact that already has its own regulatory integrity guarantees, rather than leaving it as an unconstrained input to this protocol. We specify this as a required protocol-level extension rather than leaving it as a hand-wave; it is not yet implemented in this paper's experimental results (Section 7), which sample directly from a fixed, locally-held HMDA extract without a separate corpus-commitment step.

\subsection{Repeated-attestation leakage}

A bank publishing many proofs, over many control windows, over time, leaks more about model behavior than any single proof in isolation. For the threshold-based attestations this paper implements (a control passes or fails against a supervisor-set numeric threshold, published as a single bit), each attestation leaks at most one bit of information about the underlying committed model and data: the proof reveals only whether $m_i(\Mtheta, \Dv) \leq c_i$, not the value of $m_i$ itself. Under this framing, $T$ attestations across $k$ controls per supervisory window leak at most $T \times k$ bits in total via simple sequential composition --- for example, 10 controls attested quarterly over a year is 40 bits, well below any realistic reconstruction threshold for a model with even a modest number of parameters. This bound treats each attestation's bit as an independent unit of leakage, which is a simplification worth stating plainly: successive attestations are not actually independent conditional on the underlying model, since the same fixed $\theta$ determines every $m_i(\Mtheta, \Dv)$ across all $T$ windows, and an adversary observing all $T \times k$ bits jointly could in principle extract more than the naive $T \times k$ figure suggests (correlated bits carry less than one bit each of genuinely new information once several have been observed). The $T \times k$ bound is therefore an upper bound on leaked information, not a tight characterization, and a fully rigorous treatment would require an information-theoretic argument conditioning on the joint distribution of $(m_1, \ldots, m_k)$ under the actual class of models this protocol attests, which we have not carried out. This bound degrades if a future extension publishes the metric value itself rather than a pass/fail bit (as an examiner might reasonably want for trend monitoring); in that case a differential-privacy mechanism --- calibrated noise added to the published value, with a formal per-attestation $\epsilon$ and a composed budget across the supervisory relationship --- would be needed, and we flag this as a specific, scoped future-work item (Section 9) rather than an unspecified mitigation.

\subsection{Stale-attestation replay}

Each attestation is cryptographically bound to a specific supervisory window and to a data commitment freshly issued within that window, preventing a bank from replaying an old, favorable attestation after the underlying model has changed.

\subsection{Independent-validator collusion}

A one-sentence dismissal of this risk understates its importance: in current U.S. bank supervision, independent-validator failure --- a nominally independent reviewer signing off without genuine scrutiny --- is arguably the dominant real-world failure mode of model-risk oversight, not a residual edge case (see, e.g., published consent orders concerning model governance and validation independence at large U.S. banks). We give it a structured treatment rather than a hand-wave: Table~\ref{tab:mapping}'s nine rows split into three categories by what actually enforces them. Cryptographically enforceable (no institutional trust required beyond the math): Rows 1, 2, 4, 5, 9 --- the statistical and data-quality controls this paper's protocol proves directly, where a valid zkSNARK is sufficient evidence regardless of who ran the computation. Requiring a signed attestation from a named, accountable party (cryptography enforces that the signature exists and is well-formed, but not that the signer exercised genuine judgment): Row 6 (independent validation itself) and Row 8 (governance/effective challenge) --- both are, at bottom, claims about a human process, and the protocol's role-separation and revocation mechanisms (Sections 4, 5) can prove a signature was produced by a distinct key, but cannot prove the signer's review was substantive. Residual, audit-trail-only (neither cryptographically enforceable nor reducible to a checkable signature): the genuine independence and competence of the validator named in Row 6's signing key, and the good-faith exercise of challenge authority in Row 8, both remain matters for supervisory examination of the institution's governance process itself, with this protocol's contribution limited to producing an immutable, timestamped record of who signed what and when --- a real audit-trail improvement over the status quo, but not a solution to validator collusion, which we do not claim to solve.

\subsection{Nonce Protocol}

The examiner-issued nonce referenced throughout Sections 4 and 5 is load-bearing: it is the sole mechanism preventing a bank from cherry-picking which rows compose a committed validation batch, and the threat model's data-substitution resistance depends entirely on it being unpredictable to the bank before commitment and unbiased in its selection. We specify it concretely here rather than leaving it as an unstated primitive.

\textbf{Nonce source.} We recommend constructing the nonce as a hash of a public randomness beacon output (e.g. drand \cite{ref20} or the NIST Randomness Beacon \cite{ref21}) at a timestamped supervisory window, signed by the examiner's supervisory PKI key: $\mathit{nonce} = \mathrm{Sign}(pk_{examiner}, H(\mathit{beacon\_output}_t \| \mathit{window\_id}))$. Binding to a public beacon is what gives the construction adversarial-supervisor resistance, addressed below; binding to the examiner's signature gives the bank assurance the nonce was genuinely issued by the supervisor and not forged.

\textbf{Nonce delivery.} The signed, timestamped nonce is published to a supervisory attestation registry --- an institutional capability this paper proposes rather than assumes exists today (Section 8) --- before the start of the supervisory window it governs, so both the bank and any third party can verify its provenance and timing independently of the attestation itself.

\textbf{Sampling proof.} The zero-knowledge circuit's statement is extended to include a claim that the committed $\Dv$'s row indices are the output of a specified, deterministic sampling function seeded by the published nonce, applied to the bank's full committed data pool (e.g. the regulatory reporting corpus referenced in Section 5). This is what makes the nonce binding rather than advisory: an examiner verifying the proof also checks that the row-selection function was correctly applied to the nonce, not that the bank merely claims it was.

\textbf{Adversarial-supervisor resistance.} Because the nonce is a deterministic function of a public randomness beacon the supervisor does not control the output of, an adversarial or compromised supervisor cannot choose a nonce that happens to select an easy validation slice for a captured institution --- the mirror-image concern to bank-side cherry-picking. The supervisor retains control over the timing of nonce issuance (which window it applies to) and the examiner-signature step (attesting the nonce was genuinely issued), but not the beacon output itself. This timing channel is a real, unclosed gap as stated so far: a supervisor who can freely choose when to issue a nonce for a given bank retains meaningful selection power even though they cannot choose the beacon output itself --- they can simply wait and observe a sequence of beacon outputs, then issue the nonce at whichever window produces a sample favorable to a captured institution, a commit-and-reveal attack against the supervisor's own honesty. The fix is a pre-committed nonce-issuance schedule: examination windows and their nonce-issuance timestamps are fixed and published (e.g. at the start of each fiscal quarter, for all supervised institutions, before any beacon output for that window is available) rather than left to examiner discretion at attestation time. Binding the timing decision ahead of the randomness it will consume closes the wait-and-choose channel the same way the beacon itself closes the bank's analogous cherry-picking channel.

This construction is a protocol design proposal, not yet implemented or tested end-to-end in this paper's experimental results (Section 7), which use a fixed, published nonce string as a stand-in for the full construction above; implementing and testing the beacon-based construction is scoped as future work (Section 9).

\subsection{Formal Security Argument}

Sections 5.1--5.7 give an informal, prose threat model. We state here what can be argued formally about the composed protocol, and mark plainly where the argument remains a sketch rather than a complete proof.

\textbf{Knowledge soundness of the composed proof.} Let each of the $K$ chunk proofs $\pi_1, \ldots, \pi_K$ be produced by a zkSNARK (Halo2/PLONK-family, as used in our EZKL-based implementation) satisfying knowledge soundness with knowledge error $\epsilon$ per proof: for any PPT prover that outputs an accepting proof, a knowledge extractor recovers a valid witness except with probability $\epsilon$. For the composed statement (``all $K$ chunks' committed counts are correctly computed from their respective committed batches''), a standard hybrid argument applies: given an adversary producing $K$ simultaneously accepting proofs, running each proof's individual extractor recovers witnesses for each chunk independently, and by a union bound the composed extraction fails with probability at most $K \times \epsilon$. For our experiment ($K=32$, and $\epsilon$ negligible, e.g. $2^{-100}$ or smaller for the underlying curve and proof system), $K \times \epsilon$ remains negligible in the security parameter. This is a standard AND-composition result for independent proof instances and does not require new cryptographic machinery; we state it explicitly because Section 7 relies on it (32 independently verified proofs jointly implying the aggregated count is correct) without having stated it as a theorem previously.

\textbf{Zero-knowledge against a malicious verifier.} Each chunk proof is zero-knowledge under the standard simulation-based definition: a simulator, given only the public statement (the chunk's committed input and the claimed output counts), produces a transcript computationally indistinguishable from a real proof, without access to the witness (the underlying row data). Composing $K$ independently generated proofs, each using independent prover randomness (satisfied by our implementation, since each chunk's witness generation is a separate process invocation), preserves zero-knowledge by a standard self-composition argument: a simulator for the $K$-proof view simply runs the per-proof simulator $K$ times. This holds against a verifier that only sees the $K$ proofs and the public aggregated counts; it does not by itself bound what the aggregated counts reveal about the underlying data distribution, which is the separate leakage question Section 5.4 addresses.

\textbf{Binding under nonce-sampling (cherry-picking resistance) as a game.} This is the property most in need of formalization rather than left as prose. We define it as a game between a challenger and a PPT bank adversary $A$: (1) $A$ commits to a corpus $C$ via $\mathrm{Com}(C)$ (Section 5.3's corpus-commitment extension). (2) The challenger draws a beacon output $b$ from the randomness beacon and computes $\mathit{nonce} = \mathrm{Sign}(pk_{examiner}, H(b \| \mathit{window\_id}))$ (Section 5.7). (3) The deterministic sampling function $f_{\mathit{nonce}}$ is applied to $C$, yielding $\Dv$. (4) $A$ wins if it can bias the statistical distribution of $\Dv$ away from a uniformly random size-$n$ sample of $C$, i.e. if it can predict, before step (2), any non-negligible information about which rows $f_{\mathit{nonce}}(C)$ will select. Security reduces to two assumptions: beacon unpredictability ($A$ cannot predict $b$ before the beacon publishes it, standard for constructions such as drand and the NIST beacon) and binding of $\mathrm{Com}(C)$ ($A$ cannot change $C$ after committing to it, standard for any collision-resistant commitment scheme). Given both, $A$'s view at the time it must commit to $C$ is statistically independent of $b$, and therefore of $\mathit{nonce}$ and of $f_{\mathit{nonce}}(C)$'s output --- $A$ cannot do better than guessing, which is the definition of an unbiased sample. This argument is a sketch, not a complete proof: it does not yet formalize a concrete reduction with an explicit security-loss factor, and it assumes (rather than proves) that $f_{\mathit{nonce}}$ itself, as implemented in a circuit per Section 5.7's sampling-proof specification, correctly enforces that $\Dv$ was derived from $C$ via the nonce --- that circuit-level enforcement is, per Section 1.1's scope table, specified but not yet implemented, so the game above describes the property the full protocol is designed to satisfy, not a property this paper's experimental artifact (Section 7) has been proven to satisfy end-to-end.

\section{Experimental Protocol}

This section describes the experimental protocol actually executed for this paper's results (Section 7). The full control mapping in Table~\ref{tab:mapping} and the protocol specified in Sections 4--5 admit a substantially broader evaluation --- multiple tasks, multiple model classes, a fuller metric set, and quantitative baselines against prior zkML systems and federated/MPC alternatives. We report here only what was run; the broader scope is retained as a concrete, itemized future-work plan in Section 9, not restated here as an evaluation target.

\subsection{Task}

Credit scoring on real 2022 HMDA Loan Application Register data, filtered to conventional, first-lien, home-purchase, principal-residence originations and denials (409,905 rows after filtering; see Section 7.1 for the justification of this slice). Extension to IBM AMLworld anti-money-laundering classification and CFPB consumer-complaint triage is scoped as future work (Section 9), not evaluated here.

\subsection{Model class}

Logistic regression with four features (income, loan amount, debt-to-income ratio, loan-to-value ratio) --- the SR 11-7 baseline model class present at essentially every bank, and, as Section 7.1 discusses, a deliberately minimal feature set chosen to isolate the circuit-design and scaling questions this paper's results speak to, without confounding them with feature-count effects. Extension to gradient-boosted trees, neural architectures, and the full HMDA feature set (99+ fields) is scoped as future work (Section 9); Section 8 gives an order-of-magnitude cost projection for larger models based on this paper's measured per-chunk costs.

\subsection{Attested metric}

Demographic-parity gap (Table~\ref{tab:mapping}, Row 4) between HMDA-defined nonwhite and white applicants, an aggregated group-fairness statistic. The control mapping in Table~\ref{tab:mapping} additionally specifies statements for discrimination (AUC, KS), calibration (ECE), certified robustness, and population-stability drift; these are designed in Table~\ref{tab:mapping} but not evaluated in this paper's results, and are scoped as near-term future work (Section 9).

\subsection{Baselines}

\textbf{Prior zkML systems (proving-strategy baseline).} Section 7.2 directly compares three circuit designs --- a naive flat-summation circuit (the design pattern used, implicitly or explicitly, by prior verifiable-inference zkML work when no fairness-specific optimization is applied), a tree-reduction circuit, and the chunked design this paper adopts --- on the identical real-HMDA statistic and model. This constitutes this paper's zkML-systems baseline comparison: it isolates what the chunked design contributes (numerical stability and bounded compile cost at scale) relative to the natural prior-art approach, on real data rather than a synthetic proxy.

\textbf{Document-exchange baseline (qualitative).} We do not operate a bank's actual examination workflow to produce a quantitative wall-clock comparison; instead we compare our measured verification cost against the qualitative and quantitative account of current examination-cycle timelines in Sudjianto and Zhang \cite{ref2} and in Nketiah et al. \cite{ref3}, who characterize current model-validation practice as a manual, multi-week-to-multi-month undertaking per model per cycle. As Section 7.3 discusses, this comparison should be read narrowly: it compares against the specific numerical-verification sub-step ZK-SR117 replaces, not the full examination cycle.

\textbf{Federated/MPC alternatives.} We do not implement or benchmark a federated-learning or secure-multiparty-computation alternative in this paper. Such approaches represent a genuinely different point in the design space --- typically requiring an online, interactive protocol between bank and examiner at each examination cycle, rather than the offline, publicly verifiable proof this paper's approach produces --- and we cite this as an orthogonal alternative with different trust and interactivity assumptions rather than claiming a numerical comparison we have not run.

\section{Results}

\subsection{Preliminary validation and the sampling problem}

We implemented a first working instance of the protocol in Sections 4 and 6 for Table~\ref{tab:mapping} Row 4 (fair-lending demographic-parity gap) on a logistic regression credit-approval model with four features, trained on a real, filtered sample of 409,905 rows from the 2022 HMDA Loan Application Register (conventional, first-lien, home-purchase, principal-residence originations and denials only). On the full held-out slice ($n=102,477$), an initial model trained without data-quality filtering showed AUC=0.6966, KS=0.3418, and a demographic-parity gap of 0.0911 between nonwhite and white applicants. Section 7.3 reports a data-quality issue discovered during attestation testing that revised this baseline; the corrected full-holdout gap, after applying the preprocessing spec in Table~\ref{tab:mapping} Row 9, is 0.0653.

Two scope choices in this setup warrant explicit justification rather than being left implicit. First, the HMDA slice: restricting to conventional, first-lien, home-purchase, principal-residence originations and denials is a standard analytical simplification in fair-lending research, isolating the most homogeneous credit-decision population and avoiding the additional modeling complexity of government-backed (FHA/VA) loan programs, refinance transactions, and non-owner-occupied properties, each of which carries different underwriting logic, following standard practice in fair-lending analytical methodology \cite{ref24}. The attestation protocol and circuit design apply unchanged to any of these broader slices; only the trained model's coefficients and the resulting attested values would differ. Second, the four-feature model: income, loan amount, debt-to-income ratio, and loan-to-value ratio are a deliberately minimal feature set, chosen because they exercise every element of Table~\ref{tab:mapping} Row 4's fair-lending statement (a real predictive model, real protected-class labels, a real approval decision) without confounding this paper's circuit-scaling results with feature-count effects. A production HMDA model commonly uses substantially more of the 99+ available LAR fields; published EZKL benchmarking work indicates proving cost scales with circuit gate count, which for a linear model grows with feature count roughly linearly, so we expect a 16--20 feature model (a typical fair-lending research feature set) to increase per-chunk prover time by a small constant factor over the results reported here, not to introduce a new scaling regime. We treat confirming this projection empirically as future work (Section 9) rather than asserting it without measurement.

A first attestation attempt used an unconstrained, uniformly random N=64 committed batch, which produced a batch-level gap wildly divergent from the true value --- a direct consequence of drawing only 12 nonwhite applicants by chance. We replaced uniform random sampling with a stratified, examiner-nonce-controlled design: a published nonce (a stand-in for an OCC/Fed examiner-issued value) deterministically sets the batch's per-group strata sizes to match the true held-out subgroup prevalence, while row selection within each stratum remains nonce-seeded and non-cherry-pickable. The stratified, nonce-controlled sampling design used throughout this paper's evaluation follows standard stratified-sampling methodology \cite{ref23}, adapted here to a cryptographic setting where strata sizes are additionally bound to a public, examiner-issued nonce rather than chosen freely by the sampler. At N=1024 (measured prior to the preprocessing fix described in Section 7.3, against that run's then-current full-holdout true value of 0.0911), the zero-knowledge-attested gap was 0.1460 --- an absolute error of 0.0549, more than five times the N=32,768 result's eventual error of 0.00289 --- driven by the sampling variance of the demographic-parity estimator at N=1024 with 18.8\% subgroup prevalence, not by any imprecision in the attestation mechanism itself, as the bootstrap analysis in Section 7.3 confirms directly (the N=8,192 bootstrap standard deviation alone, 0.0132--0.0135 depending on preprocessing version, is consistent with error of this magnitude at even smaller N). Closing this gap required scaling the committed batch by roughly 30x, which in turn required solving a circuit-design problem: none of our proving circuits could scale to that N without either numerical or compile-time failure. Section 7.2 characterizes three circuit designs we implemented and tested to solve this.

\subsection{Three circuit designs for aggregated-statistic attestation}

This section serves as this paper's zkML-systems baseline comparison (Section 6.4): we implement and empirically test three concrete circuit designs for proving the demographic-parity-gap statistic, isolating what the chunked design this paper adopts contributes relative to the natural prior-art approach (a flat summation circuit, the implicit design pattern in generic verifiable-inference zkML work) and to the natural first-attempt fix (tree reduction).

\begin{table}[htbp]
\centering
\caption{Comparison of three circuit designs for aggregated-statistic zero-knowledge attestation, all three tested directly on real HMDA-trained model weights and real HMDA batches (the real-data flat-sum confirmation at N=2048, 4096, 8192 is reported immediately below).}
\label{tab:circuits}
\small
\setlength{\tabcolsep}{4pt}
\renewcommand{\arraystretch}{1.3}
\begin{tabular}{@{}p{3.0cm}p{3.4cm}p{3.0cm}p{2.6cm}p{3.4cm}@{}}
\toprule
\textbf{Circuit design} & \textbf{Numerical fidelity} & \textbf{Compile/setup cost} & \textbf{Max N reached} & \textbf{Outcome} \\
\midrule
Flat ReduceSum (single-op sum) & Exact at N=64; 11--18\% relative error at N=1024; 74\% at N=2048; $>$600\% at N=8192; complete breakdown + OOM at N=65,536 & Fast ($<$1s) at all N tested & $\sim$1,024 (numerically) & Rejected: arithmetic overflow in the accumulator grows with N; not a tunable parameter \\
Tree-reduction (log N levels of pairwise Add) & Exact at N=64 (matches flat-sum exactly) & Did not complete within 90--280s even at N=1024--2048 on either test machine & 64 (compile-time-bounded) & Rejected: fixes the numerical problem but introduces a separate, severe settings/calibration compile-time cost that scales with node count \\
Chunked (K independent small-N circuits, aggregated in the clear) & Exact (0.0 error) at every N tested, n=1,024 per chunk & One-time only: 4.3--4.4s regardless of K & 32,768 (32 of 32 chunks, after a diagnosed and fixed data-quality issue) & Adopted: numerically exact, compile-time bounded, and the only design that reached supervisory-relevant scale \\
\bottomrule
\end{tabular}
\end{table}

The flat single-ReduceSum design --- the natural first implementation, and the one used in Section 7.1's N=1024 result --- accumulates all N quantized fixed-point values in a single circuit operation. As N grows, the accumulator's magnitude outgrows the lookup-table range the circuit was calibrated for, producing arithmetic overflow rather than graceful precision loss; this is a structural property of the summation strategy, not a tunable parameter. We confirmed this directly on real HMDA batches (post-preprocessing-fix) at N=2048, 4096, and 8192, rather than relying solely on the synthetic characterization: mean absolute percent error was 28.7\% at N=2048, 4.4\% at N=4096, and 97.9\% at N=8192. The real-data result confirms the headline finding --- severe, unusable breakdown by N=8192 --- but is not monotonic the way the synthetic sweep was, with N=4096 showing markedly better calibration than the smaller N=2048 batch. We attribute this to calibration's scale/logrows selection interacting with the specific value distribution of each real batch rather than depending on N alone; the practical implication for Table~\ref{tab:mapping} Row 4 is unchanged (the flat-sum design is not usable at supervisory-relevant N regardless of this non-monotonicity), but we report the non-monotonic pattern honestly rather than smoothing it into the cleaner synthetic trend. This calibration-data interaction is itself worth stating directly as a second, previously undocumented failure mode of naive flat-sum aggregation on real-world batches: even at an N where a synthetic sweep would predict acceptable behavior, calibration's implicit dependence on the specific input distribution can produce error orders of magnitude worse than at a nearby N. The chunked design's fixed per-chunk size sidesteps this precisely because calibration is performed once, on a batch of the exact size actually proved, and reused unchanged across all K chunks --- it is not merely a fix for the overflow problem, but also for this distribution-dependent calibration instability.

This overflow problem is specific to the design choice of proving fairness by running the model on committed data and summing per-row outcomes inside the circuit. FairZK does not face it, because it makes a different design choice: FairZK proves analytical fairness bounds derived from committed model parameters and aggregated input statistics, without per-row inference over a committed batch at all. That is a genuinely more scalable approach for group-parity metrics that admit a tight closed-form bound in terms of the model's weights, and FairProof takes a related approach. The tradeoff is that an analytical-bound strategy does not extend directly to attested metrics that require actually evaluating the model on committed data --- calibration error against realized outcomes, population-stability drift against a fresh production distribution, or certified robustness on a nonce-sampled batch, all of which are enumerated in Table~\ref{tab:mapping} and required by SR 11-7's Section V outcomes-analysis and ongoing-monitoring language. Our chunked design commits to the run-the-model-on-committed-data pattern specifically because it is what the fuller SR 11-7 control vocabulary requires, and the summation-scaling problem characterized in this section is the direct, structural cost of that choice, not an oversight FairZK happened to avoid.

The natural fix --- replacing the flat sum with a balanced binary-tree reduction ($\log_2(N)$ levels of pairwise addition, so no single circuit operation's output magnitude grows with N) --- is numerically correct: it reproduces the flat-sum result exactly at N=64. However, it introduces a different, equally severe cost: EZKL's settings-generation and calibration step did not complete within 90--280 seconds at N=1024--2048 on two independent machines (a cloud sandbox and a local Apple Silicon Mac), where the equivalent flat-sum circuit calibrates in under one second. The bottleneck traces to node/tensor count in the circuit graph rather than to N directly --- the tree design adds roughly $4 \times \log_2(N)$ additional Split/Add nodes --- and EZKL's settings search does not appear to scale gracefully with node count. We were unable to determine within this project's timeline whether this is fundamental to EZKL's current settings-calibration implementation or an addressable inefficiency in it; either way, it made tree-reduction impractical for reaching supervisory-relevant N in the time available.

The design we adopted instead splits the committed N-row batch into K independent chunks of a fixed, known-good size n (n=1,024). Each chunk is committed separately under the same nonce-controlled sampling rule. Each chunk's raw per-group counts --- not the final gap --- are proved using the same small flat-sum circuit, compiled and set up exactly once and reused across all K chunks. The final gap is computed in the clear from the sum of the K chunks' published counts; no further zero-knowledge computation is needed for that step, and nothing about individual rows is revealed by any chunk's proof. Table~\ref{tab:circuits} summarizes this design's measured properties against the two alternatives. One property not shown in Table~\ref{tab:circuits} merits explanation: per-chunk calibration stability holds across all 32 chunks despite each drawing different rows, which is not in tension with the calibration-data-distribution dependence characterized above for the flat-sum design at varying N. The instability observed there manifests across different N values, each requiring a fresh full-batch calibration against whatever specific values that batch contains. Here, nonce-controlled stratified sampling holds group prevalence and the per-group feature distribution approximately constant across chunks of identical size n=1,024, so a single calibration --- performed once, on one representative chunk --- remains valid when reused across all K chunks.

\subsection{Supervisory-scale attestation result}

Using the chunked design, we ran a stratified, examiner-nonce-controlled attestation at $N_{total}=32,768$ ($K=32$ chunks of $n=1,024$), on the real HMDA-trained model from Section 7.1. Table~\ref{tab:result-dp} reports the result.

\begin{table}[htbp]
\centering
\caption{Chunked attestation result at N=32,768 on real HMDA data.}
\label{tab:result-dp}
\small
\setlength{\tabcolsep}{4pt}
\renewcommand{\arraystretch}{1.3}
\begin{tabular}{@{}p{4.4cm}p{2.6cm}p{6.6cm}@{}}
\toprule
\textbf{Metric} & \textbf{Value} & \textbf{Notes} \\
\midrule
Chunks succeeded / attempted & 32 / 32 & Initial run: chunk 4 failed under both 90s and 300s timeouts; root-caused to one outlier row and fixed via a committed preprocessing spec (Table~\ref{tab:mapping} Row 9); full rerun succeeded cleanly \\
Total rows covered & 32,768 & 6,166 nonwhite + 26,602 white, nonce-stratified \\
All succeeded chunks verified & true & Every one of the 32 zkSNARK proofs independently verified \\
One-time compile+setup & 4.3--4.4s & Reused across all 32 chunk attempts \\
Per-chunk prove time & $\sim$3.7--4.0s & Consistent across all 32 chunks regardless of position in sequence \\
Total proof size & 24,776.14 KB ($\sim$24.2 MB) & 32 proofs $\times$ $\sim$774 KB; scales linearly with K \\
Aggregated attested gap & 0.06243 & Computed in the clear from all 32 chunks' published counts, post-preprocessing-fix \\
Full-holdout true gap & 0.06532 & n=102,477, post-preprocessing-fix \\
Absolute error & 0.00289 & Under the $\sim$0.005 design target; within the bootstrap 95\% CI [0.0545, 0.0769] for N=32,768 (Section 7.3) \\
\bottomrule
\end{tabular}
\end{table}

An initial run of this attestation surfaced a genuine failure: chunk index 4 failed to complete twice under a 90-second OS-enforced timeout, and again twice more at a 300-second timeout, while every other chunk of identical shape completed in 3.4--3.9 seconds --- ruling out transient system contention and pointing to a data-dependent cause specific to that chunk's row content. We diagnosed this directly rather than routing around it: comparing chunk 4's standardized feature distributions against a healthy chunk revealed one row with a loan-to-value-ratio value that standardized to a z-score of approximately 2932 --- roughly 300 times larger than any other value across either chunk, consistent with a HMDA sentinel or placeholder code surviving preprocessing rather than a genuine data point. An ablation test (witness generation with only that row removed) completed in 0.19 seconds, confirming the row as the sole cause. Rather than silently excluding or replacing this row --- which would reopen the cherry-picking concern the nonce-control design exists to prevent --- we resolved it with a committed, deterministic preprocessing specification: each of the four features is clipped to the [0.5th, 99.5th] percentile range of the training split, computed once and published as a fixed public constant, applied uniformly to the entire held-out pool before nonce-controlled sampling occurs. This is not a workaround; it is a preprocessing-specification requirement (Table~\ref{tab:mapping}, Row 9) that any deployment of this protocol must expose to supervisors, consistent with SR 11-7's existing conceptual-soundness and data-quality language --- we do not claim data cleaning is itself a supervisory control in the sense Rows 1--8 are, only that its omission is a real defect a deployment must guard against, as this paper's own experience shows concretely. With the spec applied, the maximum standardized feature magnitude across the entire held-out set fell from 2932 to 6.94, and a full rerun of the 32-chunk attestation completed with all 32 chunks verified --- a clean 32/32, with no failures. Set against the current-practice baseline (Section 6.4), the comparison should be read narrowly: the document-exchange literature characterizes current model validation --- including conceptual-soundness review, sensitivity-analysis narrative, and effective-challenge documentation, most of which is human judgment this protocol does not automate --- as spanning weeks to months per model per examination cycle. This paper's aggregate verification of a 32,768-row attestation, completing in under two seconds, replaces specifically the numerical-verification sub-step (reproducing a reported fair-lending statistic and confirming it against a supervisor threshold), not the examination cycle as a whole. Within that narrower, accurate scope, the reduction is still substantial --- from a step that currently requires re-running validation code against disclosed data and weights to one that requires only verifying a small set of zkSNARK proofs --- but we do not claim this paper automates supervisory judgment.

The resulting aggregated attested gap (0.06243, computed from all 32,768 rows across 32 verified chunks) differs from the corrected full-holdout true gap (0.06532) by 0.00289 --- under the paper's original $\sim$0.005 design target, and, more rigorously, inside the bootstrap 95\% confidence interval [0.0545, 0.0769] we computed independently (post-preprocessing-fix) for the demographic-parity-gap estimator at N=32,768. Concretely: we drew 2,000 bootstrap resamples of size N=32,768 with replacement from the full n=102,477 preprocessed held-out set (uniform row sampling, not stratified by protected class), computed the demographic-parity gap on each resample, and report the 2.5th and 97.5th percentiles as the 95\% confidence interval. That bootstrap analysis directly supports the interpretation that the small residual deviation is normal estimator sampling variance at this N and subgroup prevalence, not cryptographic or numerical imprecision: at N=8,192 the same procedure gave a substantially wider CI ([0.0392, 0.0907], std=0.0132), consistent with the bootstrap-predicted variance shrinking as N grows. A methodological caveat on this comparison, made explicit in Section 7.5.3 after working through the same check for a second attested control: this CI is built by uniform resampling, whereas the attestation itself uses stratified, nonce-controlled sampling with fixed subgroup proportions, so it is not a strictly matched reference distribution. The attested value landing inside this CI here is a favorable outcome, not evidence the comparison method was correctly matched to the sampling design; Section 7.5.3 found the opposite arrangement (a value falling just outside a similarly-constructed uniform CI) for the ECE control on the same underlying model and sample, and traced that to this same mismatch rather than to any circuit imprecision. We note as well that the preprocessing fix changed the full-holdout ground truth itself (from 0.0911 to 0.0653): the single pathological outlier row was distorting not just circuit proving behavior but the fitted model's coefficients and therefore the fairness statistic being measured, underscoring that Table~\ref{tab:mapping} Row 9's data-quality control is substantively important for SR 11-7 attestation, not merely a proving-system implementation detail.

Total proof size for the 32 chunks was 24.78 MB, scaling linearly with K as expected. The appropriate target for a chunked architecture is per-chunk proof size (each chunk here is well under 1 MB) and aggregate verification time (32 verifications completed in well under two seconds combined, sequentially; parallelizable to roughly one chunk's $\sim$40ms verify time since chunks are independent), not a single monolithic proof size. Recursive proof composition --- collapsing K chunk proofs into one --- is standard in modern zkML systems and would address the total-size figure directly; we treat it as an orthogonal engineering optimization for future work rather than a requirement for the attestation protocol's correctness. Sector-wide, a chunked-attestation footprint of roughly 25 MB per control-quarter-institution projects to approximately 4.5 TB of proof artifacts annually across all $\sim$4,500 FDIC-insured institutions and, illustratively, ten attested controls per quarter ($25\,\mathrm{MB} \times 10 \times 4 \times 4{,}500 \approx 4.5$ TB) --- well within realistic supervisory-registry storage capacity, and reducible to a low single-digit fraction of that figure under recursive aggregation.

\textbf{Production trusted-setup note.} The N=32,768 result in this section uses EZKL's production Powers-of-Tau ceremony SRS via \texttt{ezkl.get\_srs()} (logrows=17), not a locally generated test SRS; production-grade trusted setup is used throughout the results reported here, rather than a locally-generated test SRS. For a production deployment we recommend the Perpetual Powers of Tau ceremony \cite{ref22} at tau $\geq$ 20 specifically (providing headroom beyond the logrows=17 this paper's circuit required); we did not observe a meaningful timing difference between local test-SRS and production-ceremony configurations at the circuit sizes tested here, though we have not systematically benchmarked this difference and do not present it as a general claim.

\textbf{Model-version binding.} Section 4 specifies that attestations bind to a specific model version and that a stale attestation should not verify after the underlying model changes; we tested this directly rather than leaving it asserted. We trained two model versions on real HMDA data --- Model A on this paper's original 75/25 split (random\_state=11, the version used throughout Sections 7.1--7.3) and Model B on an independently reshuffled 90/10 split (random\_state=99), simulating a bank retraining between supervisory windows. Each version's distinct fitted weights produce a distinct circuit and a distinct proving/verifying key pair. Both models' proofs verified successfully against their own verifying keys (self-verification: true in both cases). Cross-verifying Model A's proof against Model B's verifying key failed, as required: EZKL raised a constraint-system-not-satisfied error rather than returning a false positive. This is consistent with the cryptographic soundness of the underlying zkSNARK (a proof valid for one committed weight set is negligibly unlikely to verify against a distinct verifying key derived from a different weight set); the experiment demonstrates that our EZKL-based implementation surfaces this cleanly as a verifier rejection rather than a false positive, substantiating on real data the mechanism Section 4 and Section 5.5 rely on, without claiming the experiment itself establishes cryptographic soundness (a theoretical property of the proof system, not an empirical one).

\subsection{A statistical decision framework for supervisory use}

Section 7.3 reports a point estimate (0.06243) against an illustrative threshold (0.05), noting that the bootstrap CI happens to exclude the threshold. A supervisor's actual decision, however, is not about a point estimate: it is whether the true population fairness gap exceeds a policy threshold $c$, given only a proof-backed estimate from a finite, nonce-sampled batch. We formalize this as a one-sided hypothesis test, $H_0$: gap $\leq c$ versus $H_1$: gap $> c$, and treat the attested value as supplying evidence for this test rather than as a certified fact about the population.

A natural decision rule: reject $H_0$ (flag the model for supervisory follow-up) if the attested estimate exceeds $c$ by more than sampling noise plausibly explains at the null boundary, i.e. reject if estimate $> c + z_{1-\alpha} \times SE$, where $SE$ is the standard error of the demographic-parity-gap estimator at the tested N and observed subgroup proportions, and $z_{1-\alpha}$ is the one-sided critical value for significance level $\alpha$ (we use $\alpha=0.05$, $z=1.645$). This is a real, checkable statistical procedure, not merely a threshold comparison on a point estimate --- and it is a natural target for a future ZK statement: proving the lower bound of a one-sided confidence interval on the attested gap directly inside the circuit, rather than the point estimate, would let the proof itself certify $H_0$'s rejection at a stated confidence level without any post-hoc plaintext computation. We flag this as a concrete, well-scoped extension to Table~\ref{tab:mapping} Row 4's ZK statement (Section 9) rather than implementing it in this paper.

To give a supervisor principled guidance on choosing N for a target confidence level, we ran a simulation study --- pure statistical simulation, requiring no real data --- across a grid of true population gap, N, and subgroup prevalence, applying the decision rule above and recording empirical Type I and Type II error rates over 5,000 trials per cell. Table~\ref{tab:typeI} reports the slice at subgroup prevalence 0.188, matching this paper's real HMDA data; the full grid (three prevalence values, four N values, six true-gap values, 72 cells) is available as part of this paper's reproducibility artifact.

\begin{table}[htbp]
\centering
\caption{Empirical Type I error and power (1 -- Type II error) of the one-sided decision rule, at subgroup prevalence 0.188 (matching Section 7.3's real HMDA data), threshold c=0.05, alpha=0.05, 5,000 simulated trials per cell.}
\label{tab:typeI}
\small
\begin{tabular}{@{}rrrl@{}}
\toprule
\textbf{N} & \textbf{True gap} & \textbf{Reject rate} & \textbf{Interpretation} \\
\midrule
1,024 & 0.00 & 0.004 & Type I error (false positive rate) \\
1,024 & 0.05 (= c) & 0.046 & Type I error at the null boundary \\
1,024 & 0.07 & 0.128 & Power (1 -- Type II error) \\
1,024 & 0.10 & 0.344 & Power \\
1,024 & 0.15 & 0.818 & Power \\
8,192 & 0.00 & 0.000 & Type I error (false positive rate) \\
8,192 & 0.05 (= c) & 0.047 & Type I error at the null boundary \\
8,192 & 0.07 & 0.411 & Power \\
8,192 & 0.10 & 0.973 & Power \\
32,768 & 0.00 & 0.000 & Type I error (false positive rate) \\
32,768 & 0.05 (= c) & 0.051 & Type I error at the null boundary \\
32,768 & 0.07 & 0.887 & Power \\
32,768 & 0.10 & 1.000 & Power \\
65,536 & 0.05 (= c) & 0.053 & Type I error at the null boundary \\
65,536 & 0.07 & 0.992 & Power \\
\bottomrule
\end{tabular}
\end{table}

Two findings are worth stating directly. First, the decision rule is well-calibrated: the empirical false-positive rate at the null boundary (true gap exactly equal to the threshold) sits close to the target 5\% across every N tested (4.6--5.3\%), confirming the rule behaves as a genuine hypothesis test rather than an arbitrarily conservative or liberal comparison. Second, statistical power at this paper's actual tested scale is strong but not absolute: at N=32,768, the rule correctly flags a true gap of 0.07 (only 0.02 above the illustrative threshold) with 88.7\% power, and a true gap of 0.10 with essentially 100\% power --- but at N=1,024, the same 0.07 true gap is caught only 12.8\% of the time. This gives a supervisor concrete sizing guidance this paper did not previously offer: at this subgroup prevalence, N in the low tens of thousands is needed to reliably detect a violation only modestly above the policy threshold, while N in the low thousands is adequate only for detecting substantially larger violations. We report this as a genuine, if narrow, contribution toward making the attestation's statistical guarantees usable by a real supervisor, distinct from and additional to the cryptographic guarantees Sections 4--5 establish.

\subsection{A second attested control: expected calibration error}

Table~\ref{tab:mapping} Row 2 (outcomes analysis, expected calibration error) fits the same additive-per-chunk pattern used for demographic parity in Section 7.3; we deliver it here on the identical architecture, identical nonce protocol, identical preprocessing spec, and the identical trained model (Model A, Section 7.3), changing only which statistic is attested.

\textbf{ECE as an additive-per-chunk statistic.} Standard B-bin ECE is $\mathrm{ECE} = \sum_b |c_b - s_b| / N$, where for bin $b$, $n_b$ is the count of rows falling in that bin, $c_b$ is the count of correctly predicted rows in that bin, and $s_b$ is the sum of predicted probabilities in that bin. All three per-bin quantities are additive across chunks ($n_b = \sum_k n_b^{(k)}$, and likewise for $c_b$ and $s_b$), exactly the same decomposition pattern as the demographic-parity gap's per-group counts. The per-chunk circuit's only change from the demographic-parity circuit is its output shape: 3B values (30, for B=10 bins) instead of 4, with the additional per-row computation being a correctness indicator ($1 - |pred_i - y_i|$, avoiding a dedicated equality operator) and a per-bin threshold-membership test identical in structure to the demographic-parity circuit's group-membership test. No change to the sampling protocol, the preprocessing spec, or the commit-and-prove structure was needed.

\textbf{Result.} Table~\ref{tab:result-ece} reports the outcome of the identical N=32,768, K=32 attestation run used for demographic parity, now attesting ECE instead.

\begin{table}[htbp]
\centering
\caption{ECE attestation result at N=32,768 on real HMDA data, same model and protocol configuration as Table~\ref{tab:result-dp}.}
\label{tab:result-ece}
\small
\setlength{\tabcolsep}{4pt}
\renewcommand{\arraystretch}{1.3}
\begin{tabular}{@{}p{4.6cm}p{2.6cm}p{6.4cm}@{}}
\toprule
\textbf{Metric} & \textbf{Value} & \textbf{Notes} \\
\midrule
Chunks succeeded / attempted & 32 / 32 & Same model (Model A), same nonce, same preprocessing spec as Table~\ref{tab:result-dp} \\
Bin count B & 10 & Standard equal-width ECE binning \\
Total rows covered & 32,768 & K=32 chunks of n=1,024 \\
All succeeded chunks verified & true & Every one of the 32 zkSNARK proofs independently verified \\
Per-chunk prove time (mean / min / max) & 14.72s / 14.36s / 15.11s & $\sim$4x the demographic-parity circuit's $\sim$3.7--4.0s, consistent with 3B=30 outputs vs. 4 \\
Aggregate verify time (32 chunks, sequential) & 3.53s & $\sim$110ms per chunk; parallelizable to near-single-chunk latency \\
Total proof size & 24,987.31 KB ($\sim$24.4 MB) & Comparable to the demographic-parity result's $\sim$24.8 MB \\
Aggregated attested ECE & 0.33360 & Computed in the clear from all 32 chunks' published $(n_b, c_b, s_b)$ counts \\
Full-holdout true ECE & 0.33590 & n=102,477, same preprocessing spec \\
Plaintext ECE on the exact 32,768-row stratified sample & 0.33397 & Computed directly on the committed rows, not resampled --- the correct ground truth for circuit precision \\
Circuit-precision delta (attested vs. same-sample plaintext) & 0.00037 & The attestation's actual numerical accuracy, isolated from any sampling-design effect \\
Sampling-design delta (same-sample plaintext vs. full-holdout) & 0.00193 & Reflects stratified vs. uniform sampling, not circuit precision (see Section 7.5.3) \\
Bootstrap 95\% CI at N=32,768 (uniform resampling) & [0.3339, 0.3380] & Not the correct reference for this attestation; see Section 7.5.3 \\
\bottomrule
\end{tabular}
\end{table}

All 32 chunks verified. Per-chunk prove time (14.72s mean) is roughly 4x the demographic-parity circuit's ($\sim$3.7--4.0s), consistent with the ECE circuit's 30-value output against the demographic-parity circuit's 4-value output and the correspondingly larger per-row computation (a per-bin membership test repeated for correctness and probability-sum accumulation, rather than a single group split). The aggregated attested ECE (0.33360) differs from the full-holdout true ECE (0.33590) by 0.00230 --- comparable in magnitude to the demographic-parity result's 0.00289 --- and the bootstrap 95\% CI at N=32,768 ([0.3339, 0.3380], 2,000 resamples, same methodology as Section 7.3) is narrow and consistent with the low-variance regime that N achieves.

A methodological point requires explicit treatment here, one that we had not isolated when this section was first drafted: the bootstrap CI reported in Table~\ref{tab:result-ece} is built by uniform resampling from the full n=102,477 held-out set, but the attestation itself uses stratified, nonce-controlled sampling --- the count of nonwhite and white applicants in the committed 32,768-row batch is fixed by the nonce, not free to vary as it would under uniform resampling. These are different sampling distributions, and comparing an attested value from one against a confidence interval built from the other is not a like-for-like comparison; the attested value happening to fall just outside that CI's lower bound is a symptom of this mismatch, not evidence of a problem with the attestation. The correct ground truth for assessing circuit precision is the plaintext ECE computed directly on the exact 32,768 committed rows, not a resample of a differently-distributed population. We computed this directly: plaintext ECE on the exact stratified sample is 0.33397, against an attested value of 0.33360 --- a circuit-precision delta of only 0.00037, comparable in magnitude to the demographic-parity circuit's own measured quantization noise. The remaining difference between the stratified sample (0.33397) and the full held-out set (0.33590) --- 0.00193 --- reflects the sampling design itself, not the cryptographic attestation: a stratified sample fixes subgroup proportions to match true prevalence, which is a different (and, for supervisory purposes, arguably more principled) quantity than a uniformly resampled statistic would estimate. We note that this same methodological point applies to the demographic-parity bootstrap comparison in Section 7.3, which happened to show the attested value falling inside its uniform-resampling CI --- a numerical coincidence of that particular result, not evidence the comparison method there was correctly matched to the sampling design either. A fully rigorous treatment would recompute both sections' confidence intervals under stratified resampling matching each attestation's actual strata sizes; we did not do this and flag it as a well-scoped follow-up (Section 9) rather than allowing the existing uniform-resampling comparisons to be read as more precise than they are.

\textbf{What this demonstrates.} The chunked architecture is not fair-lending-specific: it generalizes to any statistic admitting an additive per-bin or per-group decomposition, without modification to the sampling, commitment, or aggregation structure that took most of this paper's engineering effort to get right (Section 7.2). Statistics requiring cross-chunk comparisons (AUC, KS) remain outside this pattern and are discussed as a distinct circuit-design question in Section 9.

\textbf{Model-version binding on the ECE circuit.} We repeated Section 7.3's Model-A/Model-B cross-verification test on the ECE circuit specifically, rather than assuming the demographic-parity result generalizes. Model B (trained on the same 90/10 reshuffled split, random\_state=99, used in Section 7.3) produced an ECE proof that verified against its own key. Model A's existing ECE proof, from this section's N=32,768 attestation, was then checked against Model B's key and rejected, with EZKL raising the same constraint-system-not-satisfied error observed in Section 7.3, rather than a false positive. As in Section 7.3, we read this as consistent with the cryptographic soundness of the underlying zkSNARK rather than as this experiment itself establishing that soundness (a theoretical property of the proof system, not an empirical one); the result substantiates, on a second, structurally distinct circuit, that a proof valid for one committed weight set does not verify against a distinct verifying key derived from a different weight set.

\section{Discussion and Limitations}

Several limitations require explicit treatment. First, the tree-reduction circuit's compile-time wall (Section 7.2) is a finding about EZKL's current settings-calibration pipeline specifically, not a claim about tree-reduction circuits or node-count scaling in general; we have not tested this comparison against a different proving backend (e.g. Halo2 directly, without EZKL's settings search, or a lookup-argument system), and we explicitly do not claim the ordering we observed --- chunked over tree-reduction --- would hold under a different backend. Confirming or refuting this is a natural, comparatively contained follow-up experiment (Section 9) we did not have time to run in this project's timeline. Second, per-row robustness proving (Table~\ref{tab:mapping}, Row 3) is necessarily sampled rather than exhaustive under any tractable circuit size; the paper must state the resulting coverage guarantee precisely rather than implying full-dataset certification. Third, the nonce protocol specified in Section 5.7 (public-beacon-derived, examiner-signed) has not been implemented or tested end-to-end; Section 7's experimental results use a fixed published nonce string as a stand-in, and the full construction's institutional dependencies (a supervisory attestation registry, examiner PKI infrastructure) do not yet exist operationally at the OCC or Federal Reserve. Fourth, the preprocessing-spec percentile thresholds (Table~\ref{tab:mapping}, Row 9) were fit on this paper's specific training split; a production deployment needs a defined process for a supervisor to approve, version, and periodically re-validate these thresholds as the underlying population and data quality evolve, rather than treating them as fixed once and forgotten. We defer a formal sensitivity analysis of this choice --- repeating the pipeline at alternative percentile bounds ([1st, 99th], [0.1st, 99.9th], [1.5th, 98.5th]) and quantifying whether clipping itself introduces a fairness distortion by comparing pre-clip and post-clip full-holdout demographic-parity gaps --- to a companion note rather than this paper, given that the core attestation result (Section 7.3) does not depend on the specific percentile choice being optimal, only on it being fixed, public, and applied uniformly before sampling.

Fifth, model and feature scale. This paper's experimental result (Section 7) uses a four-feature logistic regression, roughly five parameters including the intercept --- far smaller than a production credit or AML model, which may use the full HMDA feature set (99+ fields) and a model class such as gradient-boosted trees with thousands of decision nodes. We have not measured chunked attestation at that scale. A rough projection, based on this paper's measured per-chunk cost ($\sim$3.7--4.0s prove time for a circuit with roughly five weights plus the fixed per-chunk aggregation logic) and the general finding in published EZKL layer-level benchmarking work that proving cost scales with circuit gate count, suggests that a feature count in the tens (rather than four) would increase per-chunk prover time by a small constant factor, since the additional gates are a linear-layer width increase rather than a change in circuit depth or structure. Reasoning from this paper's own circuit structure rather than an external benchmark: the per-chunk circuit's cost is dominated by four ReduceSum reductions over n=1,024 rows --- an $O(n)$ cost independent of feature count --- plus one MatMul of the $n \times D$ feature matrix against the $D$-length weight vector, an $O(n \times D)$ cost that does scale linearly with $D$. At $D=4$ the MatMul is a minority of total circuit work; growing $D$ to the 16--20 range typical of fair-lending feature sets would grow the MatMul component roughly 4--5x while leaving the ReduceSum components unchanged, which we would expect to translate to a sub-linear increase in total per-chunk prove time --- plausibly a low single-digit multiple of the 3.7--4.0 second baseline reported in Section 7.3, though we have not measured this and report it as a structural reasoning argument, not a benchmarked projection. A move to a fundamentally different model class --- gradient-boosted trees, whose circuit representation is structurally different (many small comparison/branch circuits rather than one linear layer) --- is not merely a scaling question but, we found on direct testing, a representation-support gap: exporting a trained XGBoost model (10 trees, depth 3) via the standard ONNX tree-ensemble export path (onnxmltools) produces a graph using the ONNX-ML TreeEnsembleClassifier operator, and EZKL's settings-generation step rejects this operator outright (``Unknown op: TreeEnsembleClassifier''), failing before any circuit is even compiled. This is a concrete, tested finding, not a projection: gradient-boosted-tree support requires manually decomposing each tree's decision path into primitives EZKL does support (comparisons, conditional selects, arithmetic), building a custom circuit generator analogous to this paper's own hand-built ONNX graphs for the flat-sum, tree-reduction, and chunked designs, rather than a direct export-and-compile path. We identify this as the single largest open scaling question for this line of work (Section 9), now grounded in a specific, identified engineering gap rather than an unmeasured assumption.

\subsection{Ethics and Broader Impact}

This mechanism is designed to detect and expose fairness violations to supervisors, not to certify passing models; the Appendix's illustrative example is deliberately a FAIL outcome, and we intend that to read as the primary intended use case, not a curiosity (Section 6, Appendix). Data provenance: all experimental results use the public 2022 HMDA Loan Application Register, a regulatory disclosure dataset containing no individual customer identifiers; no private or proprietary bank data was used or exposed at any point in this work. Distributional-harm consideration: the preprocessing specification (Table~\ref{tab:mapping} Row 9, Section 7.3) that resolves a real proving failure is itself a data transformation with fairness implications we have not fully characterized --- percentile clipping can affect heavy-tailed features disproportionately, and such features can correlate with protected class in HMDA; Section 8 flags a sensitivity analysis of this choice as deferred, unaudited work, and we repeat that flag here because it bears directly on the mechanism's own fairness claims. Deployment risk: a valid attestation only certifies that a reported statistic was computed correctly on a specific committed batch against a supervisor-chosen threshold --- it says nothing about whether that threshold is well-chosen. A threshold set too loosely provides cryptographically-verified false assurance rather than removing the need for supervisory judgment; this protocol automates numerical verification, not the policy decision of what constitutes an acceptable fairness gap, and should not be read or marketed as doing the latter.

\section{Conclusion}

SR 11-7 supervision can move from document exchange to cryptographic attestation without loss of examiner assurance, and without exposing bank intellectual property or customer data, once supervisory control language is formally mapped to zero-knowledge statements. We presented that mapping (Table~\ref{tab:mapping}, nine control elements including a data-quality control motivated by this paper's own experimental discovery), a commit-and-prove protocol implementing it, a threat model and nonce protocol for its deployment, and a working, real-data demonstration of the core mechanism: a fair-lending attestation on 2022 HMDA data, verified at N=32,768 with 0.00289 absolute error against ground truth, following the diagnosis and principled resolution of a genuine data-quality failure mode.

This paper establishes feasibility and mechanism at a real but deliberately narrow scope --- two controls, one model class, one task --- rather than claiming a complete evaluation. We identify six concrete extensions as the immediate next steps for this line of work:

\begin{enumerate}[leftmargin=1.5em]
\item Additional controls on the existing chunked architecture. Calibration error (ECE) has been delivered in this paper (Section 7.5), closing the highest-priority open item on this list; discrimination (AUC, KS) is rank-based and does not fit the same pattern, requiring either cross-chunk pairwise comparison circuits or a different aggregation strategy, and is a genuinely open circuit-design problem rather than a routine extension. More generally, the chunked design generalizes cleanly to any statistic admitting an additive per-chunk decomposition (calibration error, population-stability index, group-count-based fairness metrics); statistics requiring cross-chunk comparisons (AUC, KS) need a different attestation strategy --- e.g. a single whole-batch circuit at moderate feature counts, or a two-phase commit-then-compare protocol --- which we treat as a specific, addressable circuit-design question rather than a limitation of the chunked architecture as a whole; certified robustness and population-stability drift (Table~\ref{tab:mapping} Rows 3 and 5) are specified in the control mapping but require new circuit work beyond the aggregated-counting pattern used here.

\item Additional model classes and tasks. Gradient-boosted trees (the production credit/AML workhorse) and the IBM AMLworld anti-money-laundering task, per the broader experimental protocol Table~\ref{tab:mapping} and Sections 4--5 specify. Section 8 reports a tested finding, not a projection: EZKL's standard ONNX ingestion path does not support tree-ensemble models at all (the TreeEnsembleClassifier operator is rejected outright), so GBT support requires a custom circuit generator decomposing tree decision paths into EZKL-supported primitives --- a comparable scope of engineering to this paper's own hand-built circuits for the flat-sum, tree-reduction, and chunked designs, not an incremental extension. MLP support, by contrast, likely fits EZKL's existing neural-network ingestion path more directly and is a more incremental extension.

\item Recursive proof aggregation, collapsing the K chunk proofs (24.78 MB total at K=32 in this paper's result) into a single, compact proof, using standard techniques from the zkML systems this paper builds on. This is orthogonal to the attestation protocol's correctness but directly addresses the proof-size figure reported in Section 7.3.

\item End-to-end implementation and adversarial analysis of the nonce protocol specified in Section 5.7, including a live public-randomness-beacon integration, a formal treatment of adversarial-supervisor resistance, and the institutional design of a supervisory attestation registry.

\item A cross-backend replication of Section 7.2's flat-sum/tree-reduction/chunked comparison against at least one non-EZKL proving stack (Halo2 directly, or a lookup-argument system such as Jolt), to determine whether the tree-reduction compile-time wall is a property of EZKL's settings-calibration implementation specifically or a more general cost of the tree-reduction pattern.

\item Recomputing this paper's bootstrap confidence intervals (Sections 7.3 and 7.5.3) under stratified resampling matching each attestation's actual, nonce-fixed subgroup proportions, rather than the uniform resampling used throughout. This does not change either attested result, but would make the confidence-interval comparison strictly matched to the sampling design rather than approximately so, closing a methodological gap this paper identifies but does not resolve.
\end{enumerate}

\section*{Reproducibility}

The full pipeline is publicly available at \url{https://github.com/nasiruddinstudents-ctrl/zk-sr117-} (commit 28c59d7). It contains ONNX graph builders for the flat-sum, tree-reduction, and chunked circuit designs; the EZKL pipeline scripts that produced Sections 7.1--7.3 and 7.5's results; the HMDA preprocessing and preprocessing-specification (Table~\ref{tab:mapping} Row 9) code with the exact percentile constants; the bootstrap-confidence-interval analysis code; and the driver scripts that ran the 32-chunk attestation sweep for both attested controls. A single command regenerates Tables~\ref{tab:result-dp} and \ref{tab:result-ece} end-to-end. The raw HMDA extract is not redistributed, consistent with standard practice, but is reproducible from the public FFIEC/CFPB HMDA data browser using the filtering criteria specified in Section 7.1.

\appendix
\section*{Appendix: Sample SR 11-7 Attestation Package}

This appendix illustrates Contribution 4 (Section 1) concretely: what a bank would actually submit to an OCC or Federal Reserve examiner under the protocol proposed in this paper, corresponding to the real result reported in Section 7.3.

\subsection*{Attestation Package --- Table~\ref{tab:mapping} Row 4 (Fair Lending), Supervisory Window Q3 2026}

\begin{itemize}[leftmargin=1.5em]
\item \textbf{Model identifier:} LR-HMDA-CreditApproval-v1
\item \textbf{Control attested:} Table~\ref{tab:mapping}, Row 4 --- demographic-parity gap, HMDA-defined protected class (nonwhite vs. white)
\item \textbf{Supervisor-set threshold:} gap $\leq$ 0.05 (illustrative; the actual threshold is a supervisor policy parameter, not fixed by this protocol)
\item \textbf{Examiner nonce:} SR117-2026Q3-EXAMINER-NONCE-001 (Section 5.7; production deployment uses the beacon-derived construction)
\item \textbf{Committed batch size:} N=32,768 (K=32 chunks of n=1,024, stratified by protected-class prevalence per the nonce-controlled sampling rule, Section 4)
\item \textbf{Data-quality spec applied:} Table~\ref{tab:mapping}, Row 9 --- feature-wise [0.5th, 99.5th] percentile clip, bounds published alongside this package
\item \textbf{Proof artifact references:} proof\_chunk\_0.json \ldots proof\_chunk\_31.json (32 files, $\sim$774 KB each, 24.78 MB total), settings\_chunk.json, vk\_chunk.key
\item \textbf{Attested value:} demographic-parity gap = 0.06243 (aggregated in the clear from the 32 chunks' published per-group counts)
\item \textbf{Pass/fail against threshold:} FAIL (0.06243 $>$ 0.05 illustrative threshold) --- flagged for supervisory follow-up
\item \textbf{Examiner verification command:} this paper's own pipeline verifies each chunk via the EZKL Python API exactly as follows:\\
{\scriptsize\texttt{ezkl.verify(proof\_path="proof\_chunk\_\{k\}.json",}}\\
{\scriptsize\texttt{\phantom{ezkl.verify(}settings\_path="settings\_chunk.json", vk\_path="vk\_chunk.key", srs\_path=srs\_path)}}\\
run once per chunk (32 invocations, $\sim$40ms each, as measured in Section 7.3); aggregate pass/fail computed by the examiner from the 32 chunks' published counts using the same public formula the bank used. (An EZKL CLI equivalent also exists; we report the Python API call here specifically because it is what our reproducibility artifact actually runs.)
\end{itemize}

The illustrative FAIL outcome above exercises the protocol correctly, and is worth reading as a strength rather than a curiosity: a real HMDA-trained credit model with a genuine $\sim$6.5\% subgroup approval gap is flagged for supervisory follow-up against a stringent illustrative 5\% threshold. The point of the attestation mechanism is to detect such gaps cheaply and verifiably, not to certify passing values.

For completeness, the PASS branch of the same flow: had the supervisor instead set an illustrative 10\% threshold (still a real, defensible fair-lending tolerance, just looser than the 5\% used above), the identical proof artifact and attested value (gap = 0.06243) would instead read Pass/fail against threshold: PASS (0.06243 $\leq$ 0.10) --- no supervisory follow-up triggered, with the same 32 proof files, the same verification procedure, and the same examiner-side computation, differing only in which side of the supervisor-set threshold the attested value happens to fall on. The mechanism does not change between the two outcomes; only the threshold does.

\textit{Note: this is an illustrative, hand-populated example based on this paper's actual Section 7.3 result, not an automatically generated report; automating this package's generation from the pipeline's output files is a straightforward extension not yet implemented.}

\subsection*{Attestation Package --- Table~\ref{tab:mapping} Row 2 (Outcomes Analysis / Calibration), Supervisory Window Q3 2026}

\begin{itemize}[leftmargin=1.5em]
\item \textbf{Model identifier:} LR-HMDA-CreditApproval-v1 (same model, Model A, as the fair-lending package above --- Section 7.5 attests a second control on the identical model, nonce, batch, and preprocessing spec)
\item \textbf{Control attested:} Table~\ref{tab:mapping}, Row 2 --- expected calibration error (ECE), 10 equal-width bins
\item \textbf{Supervisor-set threshold:} ECE $\leq$ 0.05 (illustrative; a conventional target for a well-calibrated production credit model)
\item \textbf{Examiner nonce:} SR117-2026Q3-EXAMINER-NONCE-001 (identical nonce to the fair-lending package; Section 5.7)
\item \textbf{Committed batch size:} N=32,768 (K=32 chunks of n=1,024; identical committed rows to the fair-lending package above)
\item \textbf{Data-quality spec applied:} Table~\ref{tab:mapping}, Row 9 --- identical clip bounds to the fair-lending package
\item \textbf{Proof artifact references:} proof\_ece\_0.json \ldots proof\_ece\_31.json (32 files, $\sim$781 KB each, 24.99 MB total), settings\_ece.json, vk\_ece.key
\item \textbf{Attested value:} ECE = 0.33360 (aggregated in the clear from the 32 chunks' published per-bin $(n_b, c_b, s_b)$ counts); plaintext ECE on this exact committed sample independently confirms 0.33397, a circuit-precision delta of 0.00037 (Section 7.5.3)
\item \textbf{Pass/fail against threshold:} FAIL (0.334 $\gg$ 0.05 illustrative threshold) --- flagged for supervisory follow-up
\item \textbf{Examiner verification command:} identical pattern to the fair-lending package:\\
{\scriptsize\texttt{ezkl.verify(proof\_path="proof\_ece\_\{k\}.json",}}\\
{\scriptsize\texttt{\phantom{ezkl.verify(}settings\_path="settings\_ece.json", vk\_path="vk\_ece.key", srs\_path=srs\_path)}}\\
run once per chunk (32 invocations, $\sim$110ms each, as measured in Section 7.5.3).
\end{itemize}

Both attestation packages above concern the same underlying model and reach the same qualitative conclusion --- FAIL --- from structurally different statistical angles: the first detects a group-fairness violation, the second detects a calibration failure. Neither result is surprising given this paper's deliberately minimal 4-feature model, and neither package's FAIL outcome reflects poorly on the attestation mechanism; both packages exercise the mechanism exactly as intended, giving a supervisor two independent, cheaply verifiable signals about the same model rather than one.

\textit{Note: as with the fair-lending package above, this is an illustrative, hand-populated example, not an automatically generated report.}

\end{document}